\documentclass[aps,twocolumn,amsmath,amssymb]{revtex4}
\usepackage[dvips]{color}
\usepackage{graphicx}
\usepackage{bm}
\usepackage{xcolor}

\newcommand{\up}{\uparrow}
\newcommand{\down}{\downarrow}

\usepackage{amsmath,amssymb}
\usepackage{comment} 

\usepackage[colorlinks]{hyperref}
\hypersetup{
linkcolor = blue,
citecolor = blue,
urlcolor = black
}

\begin{document}

\title{Nodal topology in $d$-wave superconducting monolayer FeSe}

\author{Takeru Nakayama$^1$}\email{t.nakayama@issp.u-tokyo.ac.jp}
\author{Tatsuya Shishidou$^2$}
\author{Daniel F. Agterberg$^2$}
\affiliation{$^1$The Institute for Solid State Physics, The University of Tokyo, 
5-1-5 Kashiwanoha, Kashiwa, Chiba 277-8581, Japan.}
\affiliation{$^2$Department of Physics, University of Wisconsin-Milwaukee, Milwaukee, Wisconsin 53211, USA.}
\date{\today}

\begin{abstract}
A nodeless $d$-wave state is likely in superconducting monolayer FeSe on SrTiO$_3$. The lack of nodes is surprising but has been shown to be a natural consequence of the observed small interband spin-orbit coupling.  Here we examine the evolution from a nodeless state to the nodal state as this spin-orbit coupling is increased from a topological perspective. We show that this evolution depends strongly on the orbital content of the superconducting degrees of freedom. In particular, there are two $d$-wave solutions, which we call orbitally trivial and orbitally nontrivial.  In both cases, the nodes carry a $\pm 2$ topological winding number that originates from a chiral symmetry. However, the momentum space distribution of the positive and negative  charges is different for the two cases, resulting in a different evolution of these nodes  as they annihilate to form a nodeless superconductor. We further show that the orbitally trivial and orbitally nontrivial nodal states exhibit different Andreev flat band spectra at the edge. 
\end{abstract}

\maketitle
\section{Introduction}
Monolayer FeSe grown on SrTiO$_3$ has generated much attention due to its high  superconducting transition temperature $T_c$, which is higher than all the other Fe-based superconductors \cite{Wang-12}. Quasiparticle interference  \cite{Q-15} experiments and scanning tunneling microscopy~\cite{Wang-12,Zhi-14} suggest a plain $s$-wave pairing state. Angle-resolved photoemission spectroscopy (ARPES)~\cite{Defa-12, Shaolong-13, Shiyong-13, zha16} also supports this point of view by observing a fully gapped superconducting state, although with a nontrivial anisotropy \cite{zha16}. The appearance of an $s$-wave pairing state in this material seems at odds with the understanding that superconductivity in Fe-based materials is due to repulsive electron-electron interactions and presents a puzzle.  Furthermore, monolayer FeSe lacks the hole pockets about the $\Gamma$ point of the Brillouin zone (BZ) which exist in other iron pnictide compounds. This suggests that the usual 
$s_\pm$-wave pairing~\cite{Mazin-08, Kuroki-08} due to spin  fluctuations about a collinear antiferromagnetic state with a wave vector that originates from the momentum difference between electron and hole pockets is less likely as a pairing mechanism. This has led to a debate about the nature of the pairing state in monolayer FeSe. Some proposals include (for a review see Ref.~\onlinecite{Huang-17}) a conventional $s$-wave pairing state \cite{Q-15, coh15}, an incipient $s$-wave pairing state \cite{che15}, an extended $s$-wave pairing state \cite{maz11}, a fully gapped spin-triplet pairing state \cite{Eugenio-18}, and a nodeless $d$-wave pairing state \cite{li16, Agterberg-17-2}.

Recently, we revisited the nature of the magnetic correlations and the pairing state in monolayer FeSe \cite{Shishidou-17, Agterberg-17-2}.
Inelastic neutron scattering in single-crystal FeSe~\cite{Qisi-16} has found that, in addition to collinear antiferromagnetic fluctuations, there are also fluctuations associated with translation invariant checkerboard antiferromagnetic (CB-AFM) order.  First-principles spin-spiral calculations~\cite{Shishidou-17} also report the enhanced CB-AFM fluctuations in monolayer FeSe, finding that this system sits at a quantum spin-fluctuation-mediated spin paramagnetic ground state. Motivated by the presence of CB-AFM fluctuations, a symmetry-based $\bm{k}\cdot \bm{p}$ theory assuming a single $M$-point electronic representation was used to describe fermions coupled to these fluctuations \cite{Cvetkovic-13, Agterberg-17-2}. This theory 
predicts a fully gapped, nodeless $d$-wave state ~\cite{Agterberg-17-2}. Although, typically, symmetry arguments imply that such a $d$-wave state should be nodal~\cite{Sigrist-91}, this theory reveals that nodal points emerge only if the relevant interband spin-orbit coupling energy is larger than the superconducting gap.  This theory thereby naturally accounts for the gap minima that are observed along the expected nodal momentum directions of the $d$-wave state \cite{zha16}.

A natural question is, What is the mechanism that leads to a nodeless, fully gapped $d$-wave superconducting state? Indeed, one can ask how such nodeless states are more generally achieved when symmetry arguments would dictate nodes. Here we address this question through an examination of the nodal $d$-wave state.  This question falls naturally into the growing research on topological systems, 
which originally started with gapped systems~\cite{Hasan-00} such as quantum Hall systems and topological insulators in which surface states are characterized by ``bulk-edge correspondence." More recently, this was 
extended to gapless systems such as Weyl and Dirac semimetals~\cite{Vafek-14} and unconventional superconductors~\cite{Schnyder-15}. 
In unconventional superconductors that are nodal, that is, that have momenta with zero gap,
it is known that the sign change of the pairing potential on the Fermi surface leads to dispersionless Andreev bound states at a surface of the system. These states are characterized through topological arguments~\cite{Sato-11, Schnyder-12}. 
Therefore, studies of nodes in unconventional superconductors are important not only to reveal the pairing mechanism but also to clarify the topological surface states. 


Although $d$-wave superconducting states typically have topologically protected nodes in one-band systems,
these nodal points can be annihilated in multiband superconductors~\cite{Chubukov-16, Nica-17}. 
Indeed, it has been pointed out that the merging nodal points near the $\Gamma$ point have winding numbers of opposite-sign in Fe-based superconductors~\cite{Chichinadze-17}.
In addition, a nodeless $d$-wave superconductor has also been discussed in the context of cuprates~\cite{Zhu-16}. 
These works did not include spin-orbit coupling, which is essential in our theory. Our work highlights the annihilation of nodes solely due to spin-orbit coupling and demonstrates that the nodal charge is protected by a chiral symmetry that is the product of time-reversal and particle-hole symmetries. Furthermore, we find that the nodal annihilation depends upon the orbital structure of the $d$-wave gap. In particular,  we find two types of $d$-wave pairing:
(a) orbitally trivial usual $d$-wave anisotropy with a $k_x k_y$ momentum dependence and (b) orbitally nontrivial with no momentum dependence. For the latter case, nodal annihilation arises in a natural and straightforward manner, while for the orbitally trivial case, the annihilation  is much less straightforward, proceeding initially through the creation of additional nodes which then annihilate with the original nodes as the interband spin-orbit coupling is decreased.

The remainder of this paper is organized as follows. 
In Sec.~\ref{sec:model}, we introduce the symmetry-based effective model that describes the electronic excitations that stem from a single $M$ point representation of the BZ; these representations are fourfold degenerate and thus lead to two bands. We then briefly review the emergence of nodal points due to interband spin-orbit coupling. 
In Sec.~\ref{sec:Transition}, we give the topological charges for these nodal points as a $2\mathbb{Z}$ invariant 
and show that there are topologically distinguished phases which manifest themselves through the presence of dispersionless Andreev surface states.
The results are summarized in Sec.~\ref{sec:Conclusion}.


\section{Model}
\label{sec:model}
In this section, we present a brief review of the low-energy symmetry-based $\bm{k}\cdot\bm{p}$-like theory that describes the electronic states of monolayer FeSe in the vicinity of the Fermi level~\cite{Agterberg-17-2}.  
 Density functional theory calculations show that two states, which are $\bm{k}$-dependent linear combinations of Fe $\{xz, yz\}$ and $x^2-y^2$ orbitals, which are the two electronic $M$-point representations $M_1$ and $M_3$ using the nomenclature of Ref.~\cite{Cvetkovic-13}, are dominant at the Fermi level around the $M$ point. 
 These states can be described as originating from a single $M$-point four-fold electronic representation (with two orbital and two spin degrees of freedom) through an effective $\bm{k}\cdot\bm{p}$ theory. The simplicity of this model allows insight into the underlying physics that cannot be found  using a theoretical model simply based on ten orbital and two spin degrees of freedom. In addition, it captures the relevant physics of the superconducting state that appears in theories of monolayer FeSe that include two $M$-point representations \cite{Eugenio-18}.

\begin{figure}
\begin{center}
  \includegraphics[width=\linewidth]{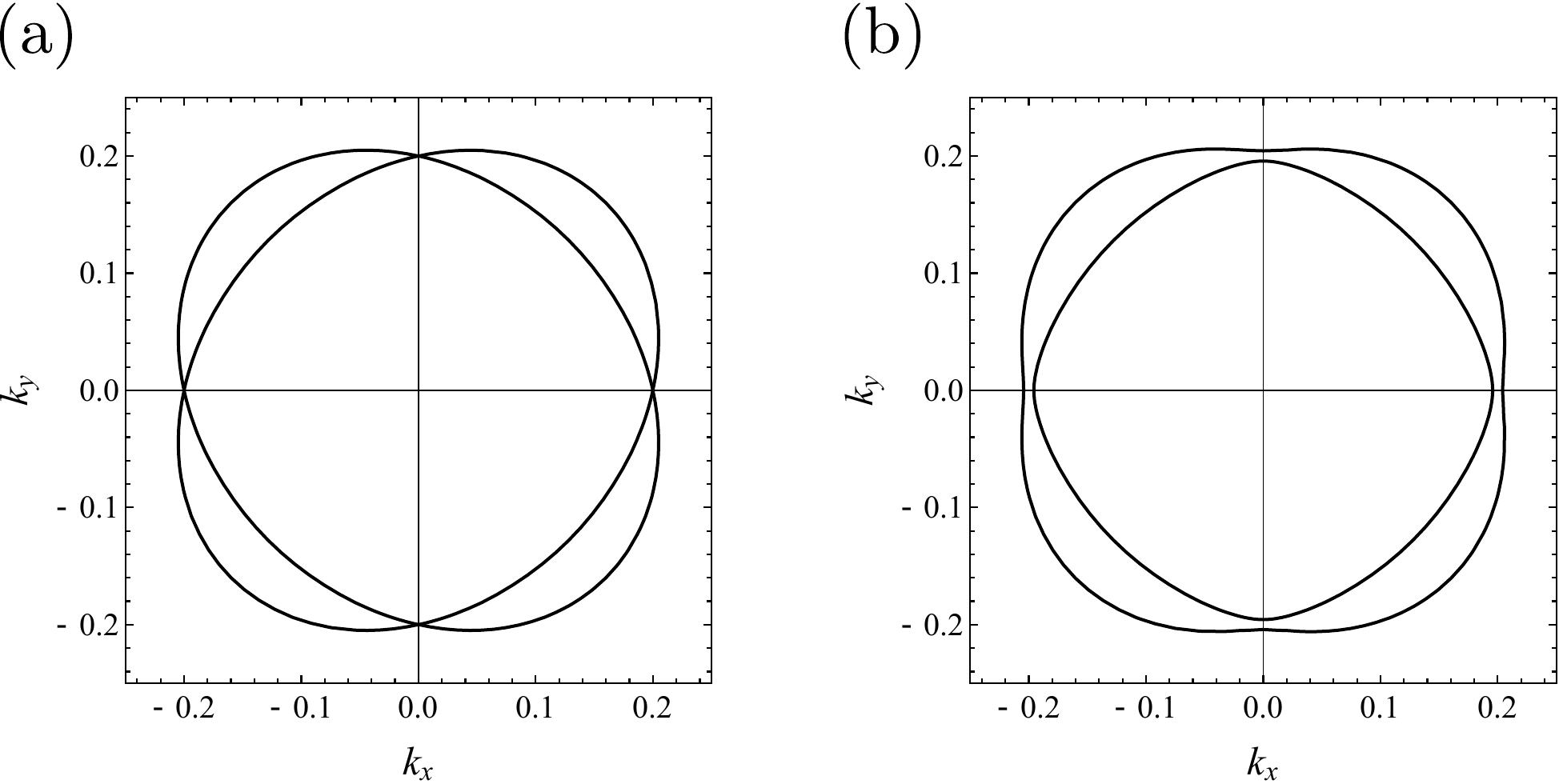}
   \caption{Fermi surfaces in normal states (a) without spin-orbit coupling and (b) with spin-orbit coupling $v_{\rm so} =12$~meV~\AA. The units of horizontal and vertical axes are \AA$^{-1}$. 
  The other parameters are given in the main text.}
 \label{fig:FermiSurface}
 \end{center}
\end{figure}

In this theory, the normal-state Hamiltonian is
\begin{equation}
H_0(\bm{k})=\epsilon_0\tau_0\sigma_0+\gamma_{xy}\tau_z\sigma_0 
+\tau_x\left[\gamma_x\sigma_y+\gamma_y\sigma_x\right],
\end{equation}
where $\bm{k}=(k_x, k_y)$ is the momentum measured from the $M$-point of the BZ and the $\tau_i$ ($\sigma_i$) matrices describe the two orbital (spin) degrees of freedom. 
The $\tau_x$ term is the interband spin-orbit coupling that plays an essential role in the $d$-wave superconducting state. This term has a magnitude that is related to the on-site spin-orbit coupling but is also determined by other factors and can be small even if the on-site spin-orbit coupling is substantial.
The Fermi surface, as observed by ARPES, is reasonably described 
when we chose  $\epsilon_0=\epsilon_0(\bm{k})=(k_x^2+k_y^2)/2m-\mu$, $\gamma_{xy}=\gamma_{xy}(\bm{k})=ak_xk_y$, $\gamma_x=\gamma_x(\bm{k})=v_{\rm so}k_x$, $\gamma_y=\gamma_y(\bm{k})=v_{\rm so}k_y$ and parameters as $\mu = 55$~meV, $1/(2m) = 1375$~meV~\AA$^2$~, $a = 600$~meV~\AA$^2$~ and $\left|v_{\rm so}\right| \le15$~meV~\AA~. 
The normal state dispersions are given by $\xi^\pm=\epsilon_0\pm\sqrt{\gamma_{x}^2+\gamma_{y}^2+\gamma_{xy}^2}$,  which have positive helicity and negative helicity, respectively.
Figures~\ref{fig:FermiSurface}(a) and \ref{fig:FermiSurface}(b) show the Fermi surfaces without spin-orbit coupling and with spin-orbit coupling $v_{\rm so} =12$~meV~\AA, respectively. 

Superconducting pairing is assumed to be induced by the fluctuations associated with translation-invariant CB-AFM.  This yields a $d_{xy}$-like pairing state. Importantly, for this paper, there are two such pairing states that are described in more detail below. The Hamiltonian is given by the following in the Bogoliubov-de Gennes form: 
\begin{eqnarray}
H(\bm{k})&=&\Gamma_z \left(\epsilon_0\tau_0\sigma_0+\gamma_{xy}\tau_z\sigma_0+\gamma_x\tau_x\sigma_y\right)\nonumber \\
&\quad&+\gamma_y\Gamma_0\tau_x\sigma_x+i\Gamma_y\left(\Delta_{d,0}\tau_0+\Delta_{d,z}\tau_z\right)i\sigma_y,
\label{eq:Hbdg}
\end{eqnarray}
where the $\Gamma_i$ matrices describe the particle-hole degree of freedom,
\begin{eqnarray}
\begin{split} 
\Delta_{d,0}&=\Delta_{d,0}(\bm{k})=\Delta_2k_xk_y/k_0^2,\\
\Delta_{d,z}&=\Delta_{d,z}(\bm{k})=\Delta_0,
\end{split}
\label{eq:delta}
\end{eqnarray}
and we take the typical Fermi wave vector $k_0=0.2$~\AA$^{-1}$. The two gap functions $\Delta_{d,0}$ and $\Delta_{d,z}$ are the two $d_{xy}$ pairing degrees of freedom mentioned above. 
The pairing term $\Delta_{d,0}\tau_0$ represents an orbitally trivial and usual $d_{xy}$ pairing with a $k_x k_y$ momentum dependence.  $\Delta_{d,z}\tau_z$ represents an orbitally nontrivial pairing state with no momentum dependence;  it also has $d_{xy}$ pairing symmetry due to the $\tau_z$ orbital dependence and the different symmetries of the two orbitals that give rise to this gap function. In general, since both $\Delta_{d,0}$ and $\Delta_{d,z}$ channels have the same symmetry, the gap function will be a linear combination of both these pairing channels. 

\begin{figure}
\begin{center}
  \includegraphics[width=\linewidth]{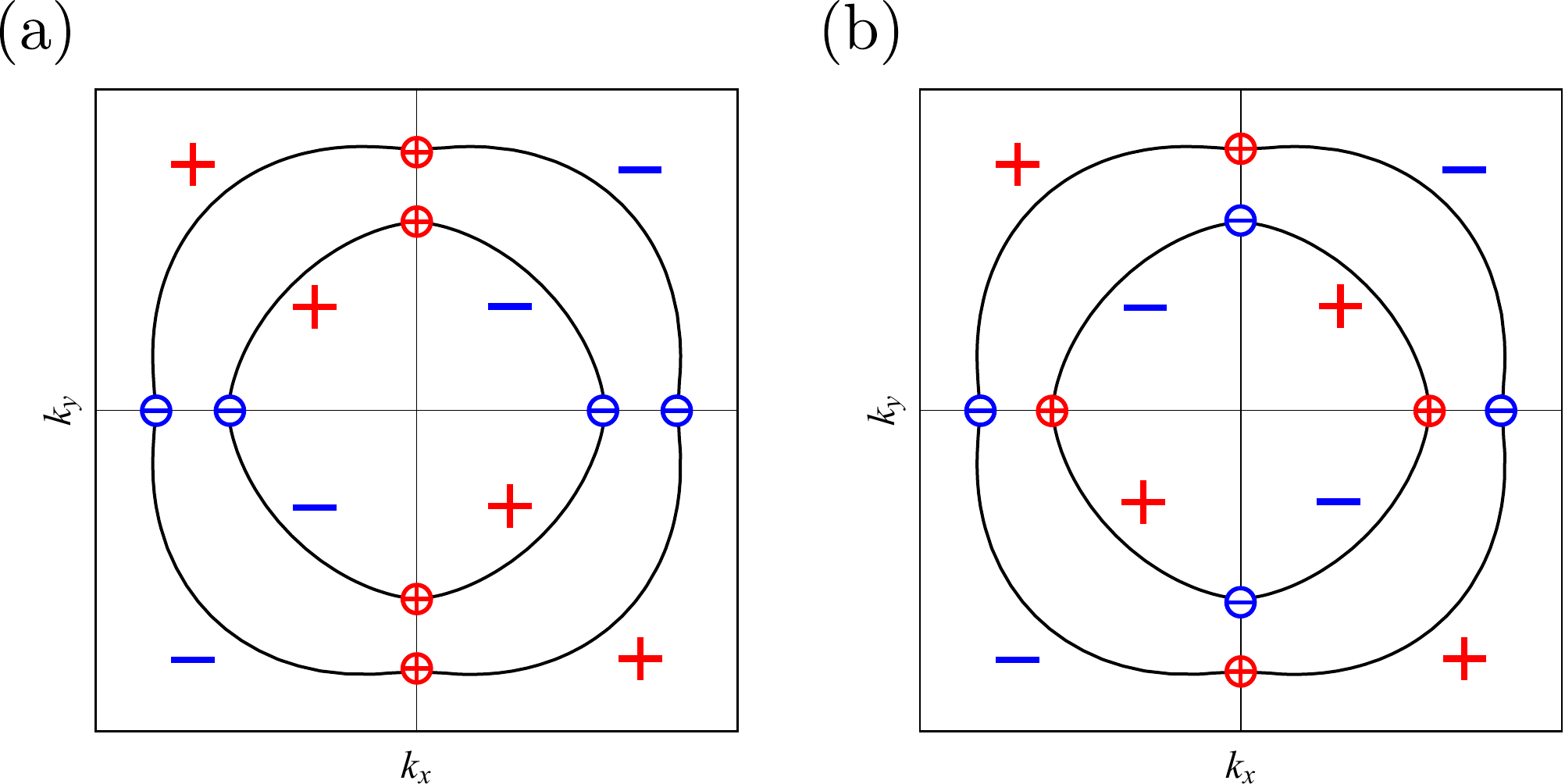}
   \caption{Pairing anisotropy and topological charges in (a) orbitally trivial pairing and (b) orbitally nontrivial pairing in the band basis with only intraband pairing. 
   The solid lines represent the Fermi surface in normal states. The circles represent $\pm2$ topological charge.}
 \label{fig:trivial_nontrivial}
 \end{center}
\end{figure}

\begin{widetext}
In order to gain a deeper understanding of these two types of $d_{xy}$ order, it is  convenient to change basis from the orbital basis to the band  basis. 
The Hamiltonian in~(\ref{eq:Hbdg}) can be written in block diagonal form with two $4\times4$ matrices. One of these matrices is
\begin{eqnarray}
\left[
    \begin{array}{cccc}
      \epsilon_0+\gamma_{xy} & \gamma_y-i\gamma_x & 0 &\Delta_{d,0}+\Delta_{d,z} \\
       \gamma_y+i\gamma_x & \epsilon_0-\gamma_{xy} & -\Delta_{d,0}+\Delta_{d,z}& 0 \\
       0 & -\Delta_{d,0}+\Delta_{d,z} & -\epsilon_0+\gamma_{xy} &   \gamma_y+i\gamma_x\\
       \Delta_{d,0}+\Delta_{d,z} & 0 & \gamma_y-i\gamma_x  & -\epsilon_0-\gamma_{xy} 
    \end{array}
  \right], 
\end{eqnarray}
while the other matrix is given by transforming  $\Delta_i\to -\Delta_i$ and $\gamma_x\to -\gamma_x$.
Performing a unitary transformation that diagonalizes the normal part of the Hamiltonian we obtain in the band basis, we find
\begin{eqnarray}
\left[
    \begin{array}{cccc}
      \epsilon_0+\sqrt{\gamma_x^2+\gamma_y^2+\gamma_{xy}^2} & \Delta_{d,0}+\frac{\Delta_{d,z}\gamma_{xy}}{\sqrt{\gamma_x^2+\gamma_y^2+\gamma_{xy}^2}}&0& \frac{\Delta_{d,z}\left(\gamma_y-i\gamma_x\right)}{\sqrt{\gamma_x^2+\gamma_y^2+\gamma_{xy}^2}}  \\      
       \Delta_{d,0}+\frac{\Delta_{d,z}\gamma_{xy}}{\sqrt{\gamma_x^2+\gamma_y^2+\gamma_{xy}^2}} & -\epsilon_0-\sqrt{\gamma_x^2+\gamma_y^2+\gamma_{xy}^2} & \frac{\Delta_{d,z}\left(\gamma_y-i\gamma_x\right)}{\sqrt{\gamma_x^2+\gamma_y^2+\gamma_{xy}^2}}& 0 \\
        0&  \frac{\Delta_{d,z}\left(\gamma_y+i\gamma_x\right)}{\sqrt{\gamma_x^2+\gamma_y^2+\gamma_{xy}^2}}   & \epsilon_0-\sqrt{\gamma_x^2+\gamma_y^2+\gamma_{xy}^2}&   \Delta_{d,0}-\frac{\Delta_{d,z}\gamma_{xy}}{\sqrt{\gamma_x^2+\gamma_y^2+\gamma_{xy}^2}}\\
         \frac{\Delta_{d,z}\left(\gamma_y+i\gamma_x\right)}{\sqrt{\gamma_x^2+\gamma_y^2+\gamma_{xy}^2}}&0 &\Delta_{d,0}-\frac{\Delta_{d,z}\gamma_{xy}}{\sqrt{\gamma_x^2+\gamma_y^2+\gamma_{xy}^2}} & -\epsilon_0 +\sqrt{\gamma_x^2+\gamma_y^2+\gamma_{xy}^2} 
    \end{array}
  \right].
\end{eqnarray}
This band basis clarifies that the Hamiltonian has both intraband and interband pairings, as is the case in other proposals for nodeless $d$-wave superconductors~\cite{Nica-17}. The interband pairing arises only from the orbitally nontrivial $\Delta_{d,z}$ (in combination with the interband spin-orbit coupling). The intraband pairing contains both pairing channels. In this case,  the orbitally nontrivial $\Delta_{d,z}$ channel explicitly gains $d$-wave momentum anisotropy through the $\gamma_{xy}$ normal state term. 
Figure~\ref{fig:trivial_nontrivial} shows the pairing anisotropy in the case of only orbitally trivial pairing [Fig.~\ref{fig:trivial_nontrivial}(a)] and the orbitally nontrivial one in the band basis~[Fig.~\ref{fig:trivial_nontrivial}(b)].
Note that here only spin-singlet pairing is considered. In general, there can be  mixing of spin-singlet and -triplet pairings due to the interband spin-orbit coupling. 

The interband pairing in the band basis is essential to generate a gapless superconducting $d_{xy}$ state, provided the interband spin-orbit coupling is sufficiently small. To understand how a large interband spin-orbit coupling gives rise to nodal points, it is useful to consider 
the quasiparticle dispersion for Hamiltonian~(\ref{eq:Hbdg}). This is given by 
\begin{eqnarray}
E_\pm(\bm{k})=\sqrt{\epsilon_0^2+\gamma_{xy}^2+\gamma_x^2+\gamma_y^2+\Delta_{d,0}^2+\Delta_{d,z}^2\pm2\sqrt{\left(\epsilon_0\gamma_{xy}+\Delta_{d,0}\Delta_{d,z}\right)^2+\left(\gamma_x^2+\gamma_y^2\right)\left(\epsilon_0^2+\Delta_{d,z}^2\right)}}.
\label{eq:dispersion}
\end{eqnarray}
\end{widetext}
Notice that there are also two negative quasiparticle dispersion $-E_{\pm}(\bm{k})$ due to chiral symmetry.
Along the nodal direction $k_y=0$, so that 
 $\gamma_{xy}=\gamma_y=\Delta_{d,0}=0$, yielding $E_\pm(\bm{k})=\left|\sqrt{\epsilon_0^2+\Delta_{d,z}^2}\pm\left|\gamma_x\right|\right|$. 
Therefore, 
the following equation must be satisfied at the nodal points (labeled $\bm{k}^*$):
\begin{eqnarray}
\epsilon_0^2=\gamma_x^2-\Delta_{d,z}^2.
\label{eq:gapless_condition}
\end{eqnarray}
This means that once the interband spin-orbit coupling satisfies $|\gamma_x|>\Delta_{d,z}$, nodal points exist. As the interband spin-orbit coupling is reduced, there is  consequently a transition from a nodal $d_{xy}$ state to a fully gapped $d_{xy}$ state, which is the focus of the remainder of this paper. Note that a generic consequence of this theory is that gap minima in the fully gapped state are along the nodal directions; this agrees with what is observed in ARPES measurements.  

\section{Nodal topological charges and Andreev flat band states}
\label{sec:Transition}
\subsection{Nodal topological charges}

Now we examine how the fully  gapped $d_{xy}$ state appears as the interband spin-orbit coupling is reduced. In particular, for sufficiently large interband spin-orbit coupling we have a nodal $d_{xy}$ state, and we examine the topological charge of the nodal points. We show that topological charge at the nodal points can be defined as a $2\mathbb{Z}$ invariant. The key symmetries in defining this charge are 
time reversal (with operator $T$) and particle-hole conjugation (with operator $C$). These act on $H(\bm{k})$ as
\begin{eqnarray}
TH(\bm{k})T^{-1}=H(-\bm{k}),\\
CH(\bm{k})C^{-1}=-H(-\bm{k}),
\end{eqnarray}
where $T=K\Gamma_0\tau_0\left(i\sigma_y\right)$, $C=K\Gamma_x\tau_0\sigma_0$, and $K$ is the complex conjugate operator.
Since $T^2=-1$ and $C^2=1$, this Hamiltonian belongs to Altland-Zirnbauer class DIII~\cite{Schnyder-08}. Furthermore, we define a chiral operator $S$,
\begin{eqnarray}
S=-iTC=\Gamma_x\tau_0\sigma_y.
\end{eqnarray}
Since chiral symmetry is preserved and $S$ anticommutes with $H(\bm{k})$,  $H(\bm{k})$ can be written in block off-diagonal form using the basis in which $S$ is diagonal:
 \begin{eqnarray}
 H(\bm{k})\to V H(\bm{k})V^\dag=
 \left[
    \begin{array}{cc}
      0& q(\bm{k})\\
      q^\dag(\bm{k}) & 0\\
      \end{array}
\right],
\label{eq:diagS}
\end{eqnarray}
where
\begin{eqnarray}
q(\bm{k})&=&\epsilon_0\tau_0\sigma_0+\gamma_{xy}\tau_z\sigma_0+\gamma_x\tau_x\sigma_y+\gamma_y\tau_x\sigma_x \nonumber \\ 
&\quad&+i\left(\Delta_{d,0}\tau_0+\Delta_{d,z}\tau_z\right)\sigma_0
\end{eqnarray}
and \begin{eqnarray}
V=\frac{1}{\sqrt{2}}
\left[
    \begin{array}{cc}
    \mathbb{I}& -\tau_0\sigma_y \\
    \mathbb{I}& \tau_0\sigma_y
     \end{array}
\right],
\end{eqnarray}
where $\mathbb{I}=\tau_0\sigma_0$ is a $4\times4$ unit matrix.
Note that ${\rm det}\;q(\bm{k^*})=0$ because of the nodal condition $E_{-}({\bm{k}^*})=0$. In addition, given that chiral symmetry leads to the topological protection discussed here, we mention physically relevant perturbations that preserve and break this symmetry. In particular, the mirror glide plane symmetry-breaking term $M_I=\lambda_I\left(k_x^2-k_y^2\right)\tau_x\sigma_0$ and nematic order $\eta_Q\tau_z\sigma_0$ preserve chiral symmetry, but a Zeeman field $\bm{h}\tau_0\cdot\bm{\sigma}$ does not.

In class DIII, a topological charge can be defined by the winding number~\cite{Beri-10}, which is given by
\begin{eqnarray}
W_{\mathcal{L}}=\frac{1}{2\pi i}\oint_{\mathcal{L}}\; d k_l \; {\rm Tr} \left[q^{-1}(\bm{k})\partial_{k_l}q(\bm{k})\right],
\label{eq:topologicalcharge}
\end{eqnarray}
where the contour $\mathcal{L}$ is a loop around the nodal point. This charge is an integer $\mathbb{Z}$ invariant. In the problem we are considering, we also have parity symmetry, which ensures a twofold degeneracy of the nodal point. Consequently, the nodes have a $2\mathbb{Z}$ topological charge~\cite{Bzdusek-17}. We find that the orbitally trivial and orbitally nontrivial gap functions exhibit different nodal charge distributions in momentum space and that a topological transition exists between these two cases. 

To understand the different nodal charge distributions between the orbitally trivial and nontrivial cases (see Fig.~\ref{fig:trivial_nontrivial}), it is useful to consider the limit in which the interband pairing can be ignored. This can be achieved in the orbitally trivial case by setting $\Delta_{d,z}=0$ and in the orbitally nontrivial case by setting $\Delta_{d,0}=0$ and also requiring that the interband spin-orbit coupling satisfy $|\gamma_i|\ll |\gamma_{xy}|$. When the interband pairing can be ignored, we can consider the nodal points in each band independently.
In this case, following Ref.s~\cite{Sato-11, Schnyder-12}, Eq.~(\ref{eq:topologicalcharge}) can be simplified to 
\begin{eqnarray}
W_{\mathcal{L^{\pm}}}=-\sum_{\bm{k}_0 \in S_{\mathcal{L}^\pm}} {\rm sgn}\left(\left.\partial_{k_{l}} \xi^\pm_{\bm{k}}\right|_{\bm{k}=\bm{k}_0} \right) {\rm sgn} \left(\Delta^\pm_{\bm{k}_0}\right),
\end{eqnarray}
where $\xi^\pm=\epsilon_0\pm\sqrt{\gamma_{x}^2+\gamma_{y}^2+\gamma_{xy}^2} $, $\Delta^\pm_{\bm{k}}$ is the superconducting gap of positive and negative helicity,
 and the sum is over the set of points $S_{\mathcal{L}^\pm}$ given by the intersection of positive- and negative-helicity Fermi surfaces with the one-dimensional contour $\mathcal{L}^\pm$. 
We consider explicitly the topological charges of the adjacent pair of nodal points in the $k_x(>0)$ direction, $(k_x^{*-},0)$ and $(k_x^{*+},0)$. 
In the orbitally trivial case,  the superconducting gap $\Delta_{\bm{k}}^\pm$ of each band is $\Delta^\pm_{\bm{k}}=-\Delta_{d,0}$. 
Therefore, two nodal points will have same-sign topological charge, which we call  {\it same-sign pair} states.
On the other hand, for the orbitally nontrivial case, $\Delta_{\bm{k}}^\pm \sim \mp\gamma_{xy} \Delta_{d,z}$, 
so that the two nodal points have opposite-sign topological charges, which we call {\it opposite-sign pair} states. In general, the pairing state will be a linear combination of the orbitally trivial and orbitally nontrivial gap functions, but it is intuitively clear that the nodes can still be classified as same-sign pair or  opposite-sign pair states and a transition between these two topological states can occur. Furthermore, in both cases, as the spin-orbit coupling is decreased, a gapped $d_{xy}$ superconducting state must arise (assuming that $\Delta_{d,z}\ne 0$). The development of this gapped state for opposite-sign pair states is intuitively clear, but this is not the case for same-sign pair states.


To gain a deeper understanding of the physics discussed above, we consider a more general treatment of the topological charge. In particular, 
the topological charge~(\ref{eq:topologicalcharge}) can be cast in the following form:
\begin{eqnarray}
\scalebox{0.9}{$\displaystyle
W_{\mathcal{L}}= \frac{1}{\pi} \oint_{\mathcal{L}}\; dk_l\; \partial_{\bm{k}} \tan^{-1}\left[\frac{2\left(\epsilon_0\Delta_{d,0}-\gamma_{xy}\Delta_{d,z}\right)}{\epsilon_0^2-\gamma_x^2-\gamma_y^2-\gamma_{xy}^2-\Delta_{d,0}^2+\Delta_{d,z}^2}\right].
$}\nonumber\\
\end{eqnarray}
This can be understood  as the winding number of the vector $(\epsilon_0^2-\gamma_x^2-\gamma_y^2-\gamma_{xy}^2-\Delta_{d,0}^2+\Delta_{d,z}^2,\;\epsilon_0\Delta_{d,0}-\gamma_{xy}\Delta_{d,z})$ rotating around the nodal point. 
The crucial term which determines whether same- or opposite-sign pairs appear is the numerator $\epsilon_0\Delta_{d,0}-\gamma_{xy}\Delta_{d,z}$ (the denominator $\epsilon_0^2-\gamma_x^2-\gamma_y^2-\gamma_{xy}^2-\Delta_{d,0}^2+\Delta_{d,z}^2$ behaves similarly for both same and opposite-sign pairs).
Substituting detailed forms~(\ref{eq:delta}), the numerator is given by
\begin{eqnarray}
\epsilon_0\Delta_{d,0}-\gamma_{xy}\Delta_{d,z}=\left\{ \begin{array}{ll}
    -ak_xk_y\Delta_0 & \Delta_2=0, \\
    \\
    \frac{k_xk_y}{k_0^2}\Delta_2\left[\epsilon_0-ak_0^2\frac{\Delta_0}{\Delta_2}\right] & \Delta_2\neq 0.
  \end{array} \right. \nonumber 
  \\
\end{eqnarray}
If $\Delta_2=0$, the sign of the numerator is the same between the two nodal points $\bm{k}^{*-}$ and $\bm{k}^{*+}$, leading to topological charges of opposite signs at the two nodal points, that is, opposite-sign pair states.  
However, if $\Delta_2\neq0$ and the sign of $\epsilon_0-ak_0^2\Delta_0/\Delta_2$ changes between the two  nodal point $\bm{k}^{*-}$ and $\bm{k}^{*+}$,  the topological charges have the same sign at the two nodal points, leading to same-sign pair states. 
In order to develop an analytic condition to distinguish these two cases, we consider the $k_y=0$ direction and set $\tilde{k}_x$ as $\epsilon_0(\tilde{k}_x)-ak_0^2\Delta_0/\Delta_2=0$. 
In the case of same-sign pair states, $k_x^{*-}<\tilde{k}_x<k_x^{*+}$, this is not satisfied for opposite-sign pair states. 
With the nodal condition~(\ref{eq:gapless_condition}), we get the following inequality:
\begin{eqnarray}
2mv_{\rm so}^2-m\sqrt{2\frac{\mu}{m}v_{\rm so}^2-\frac{\Delta_0^2}{m^2}+v_{\rm so}^4}< a\frac{\Delta_0}{\Delta_2}k_0^2\nonumber \\ 
<2mv_{\rm so}^2+m\sqrt{2\frac{\mu}{m}v_{\rm so}^2-\frac{\Delta_0^2}{m^2}+v_{\rm so}^4}.
\end{eqnarray}
As an example, if we take the values $\Delta_0=11$~meV and $\Delta_2=-1.5$ meV, which were used earlier to generate a gap anisotropy consistent with experiment, 
and assume a strong interband spin-orbit coupling $v_{\rm so}= 80$~meV \AA, then the topological character of nodal points is classified as opposite-sign pair states. 

\begin{figure}
\begin{center}
  \includegraphics[width=\linewidth]{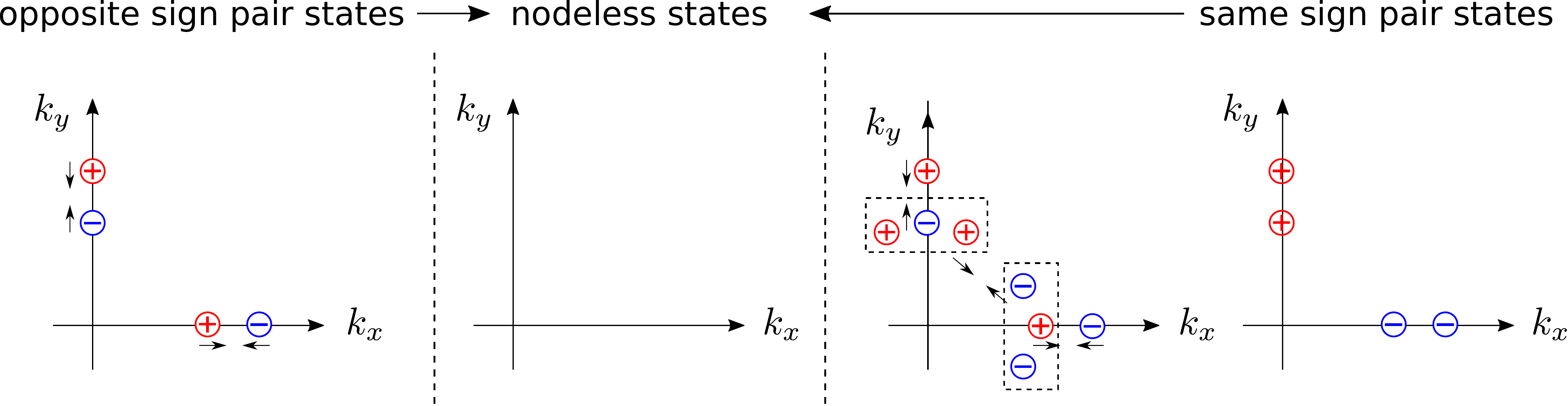}
   \caption{Schematic picture of transition to nodeless states from {\it opposite}- (left) and {\it same-sign pair} states (right). 
   The arrows represent that two nodal points merge with each other.
   In same-sign pair states, each inner nodal point splits into three nodal points (surrounded by a dashed line) in transition to nodeless states.}
 \label{fig:oppsite_to_same}
 \end{center}
\end{figure}

Now we turn to the development of the gapless $d_{xy}$ state due to the merging and annihilation of nodal points. 
It is worth emphasizing that this has been studied in Dirac and Weyl semimetals~\cite{Vafek-14} and also in $s$ and $d$-wave superconductors~\cite{Chichinadze-17} in a framework different from ours in which spin-orbit coupling is not an essential interaction. 
In the case of opposite-sign pair states, the nodal points can merge and are annihilated as the interband  spin-orbit coupling decreases because they have opposite topological charges. 
However,  in the case of same-sign pair states, merging and annihilation of nodal points cannot occur directly. We find that  this annihilation occurs through an involved mechanism. 
Indeed, as the interband spin-orbit coupling is decreased from the same-sign pair state (which we take to be positive for both in the description that follows), a new pair of opposite-charge nodal points is created near the nodal point at $\bm{k}^{*-}$. As the interband spin-orbit coupling is further decreased, the negatively charged nodal point stays near  $\bm{k}^{*-}$, while the two positively charged nodal points move off the $k_x$ (or $k_y$) axis. The positively charged nodal points that move off the $k_x$ axis  eventually merge  with similarly formed negatively charged nodal points that have moved off the $k_y$ axis. This leaves an opposite-sign pair state, for which the nodes merge and annihilate as before when the interband spin-orbit coupling is further decreased  
(see Fig.~\ref{fig:oppsite_to_same}). 


\subsection{Andreev flat-band states}

We find that, typically, either same-sign pair states or opposite-sign pair states occur when the superconducting state is nodal. In particular, the state we find above with 16 nodal points exists only in a narrow range of parameters, so we do not consider it further here. It would be of interest to be able to experimentally identify whether same-sign or opposite-sign pair states exist. As we show below, this can be done through an examination of edge states. Prior to discussing this, we note that the values of the spin-orbit coupling used below are larger than those observed in monolayer FeSe grown on SrTiO$_3$. Consequently, we do not predict flat-band energy states for this material  (however, there still exist in-gap edge states that are not topologically protected). In this context we note that the spin-orbit coupling may be larger when monolayer FeSe is grown on a different substrate or if it is doped, for example, with Te, which may allow for the flat-band edge states to be observed.

\begin{figure}
\begin{center}
   \includegraphics[width=\linewidth]{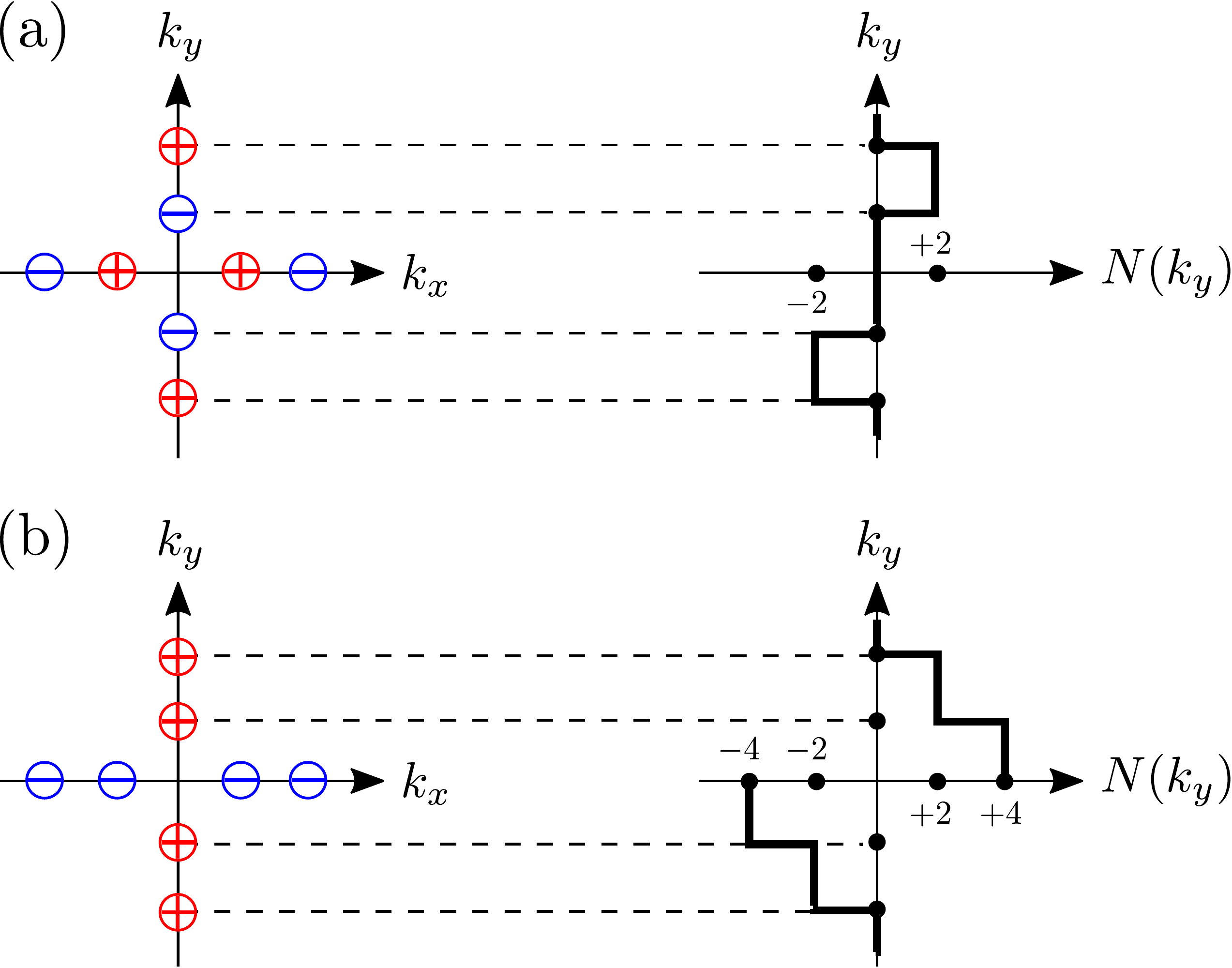}
   \caption{Schematic pictures of the relation between $W_{\mathcal{L}}$ (left) and $N(k_y)$ (right) in the case of (a) opposite-sign pair and (b) same-sign pair states. Red and blue points indicate $W_{\mathcal{L}}=+2$ and $-2$, respectively. }
 \label{fig:windingABS}
 \end{center}
\end{figure}

The nontrivial topological charges at nodal points imply the existence of dispersionless Andreev band states or Andreev flat band states as edge states. 
The number of Andreev flat-band states is related to a one-dimensional (1D) winding number $N(\bm{k}_{\parallel})$\cite{Schnyder-11, Sato-11}, which is given by
\begin{eqnarray}
N\left(\bm{k}_{\parallel}\right)= \int\; d\bm{k}_{\perp}\; {\rm Tr} \left[q^{-1}(\bm{k})\partial_{\bm{k}_{\perp}}q(\bm{k})\right],
\label{eq:1Dwinding}
\end{eqnarray}
where $\bm{k}_{\parallel}$ $(\bm{k}_{\perp})$ is the bulk momentum parallel (perpendicular) to the surface. 
We consider edges running along the $y$ direction and take $\bm{k}_{\parallel}$ ($\bm{k}_{\perp}$) as $k_y$ $(k_x)$. 
Figure~\ref{fig:windingABS} shows the relation between the 1D winding number $N(k_y)$ and the topological charge $W_{\mathcal{L}}$.
Figure~\ref{fig:windingABS}(a) shows the 1D winding number is nonzero between nodal points which have opposite-sign topological charges  but is zero at the origin in the case of opposite-sign pair states. 
On the other hand, the 1D winding number is nonzero for all momenta between the outer  nodal points in the case of same-sign pair states [Fig.~\ref{fig:windingABS}(b)].

\begin{figure}
\begin{center}
   \includegraphics[width=\linewidth]{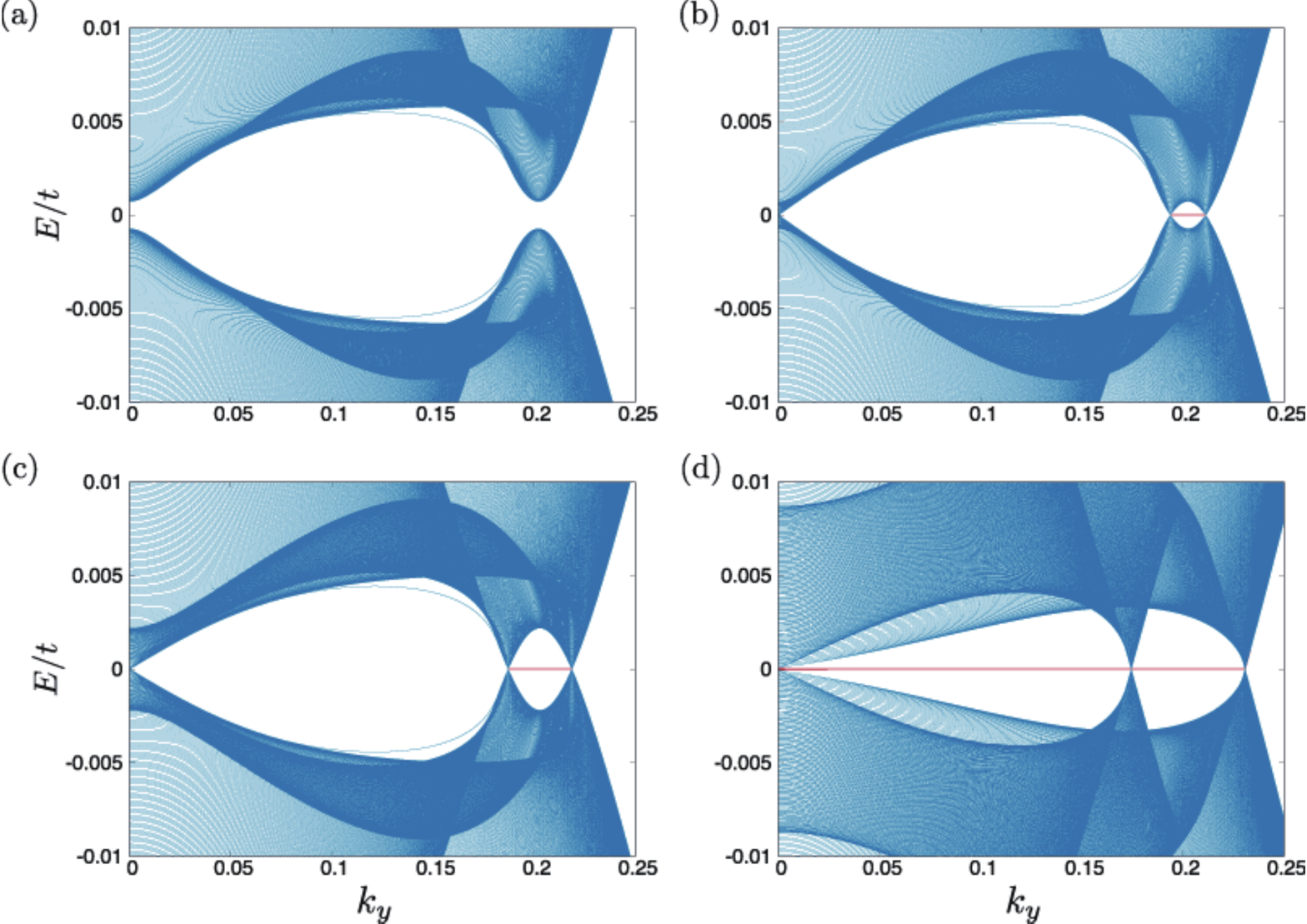}
   \caption{Energy spectra for (a) no nodal points and (b) opposite-sign pair, (c) opposite-sign pair, and (d) same-sign pair states. 
   We set the parameters  $(v_{\rm so}{\rm [meV}\mathrm{\mathring{A}}], \Delta_0 {\rm [meV]}, \Delta_2 {\rm [meV]})$ as (a) $(50, 11, -1.5),$ (b) $(60, 11, -1.5),$ (c) $(70, 11, -1.5)$, and (d) $(80, 4, -10)$. 
   The vertical axis is scaled by $t=(2m)^{-1}$. }
 \label{fig:band}
 \end{center}
\end{figure}

In order to investigate the edge states further, we introduce a lattice model which corresponds to Eq.~(\ref{eq:Hbdg}) (see Appendix~\ref{appendix:latticemodel}). 
We suppose that the system has two edges at $i_x =1$ and $N_x$ in the $x$ direction and take the boundary condition in the $y$ direction to be periodic. 
Then, we examine the edge states by numerically obtaining the energy spectrum as a function of the momentum $k_y$. We set $N_x=10000$.
Figure~\ref{fig:band} shows the energy spectra for no nodal points [Fig.~\ref{fig:band}(a)], opposite-sign pair states [Fig.~\ref{fig:band}(a) and Fig.~\ref{fig:band}(c)], and same-sign pair states [Fig.~\ref{fig:band}(d)].
Indeed, with no nodal points we do not have Andreev flat-band states, and
once the nodal points appear with increasing interband spin-orbit coupling, flat-band states appear. 
In the cases of opposite-sign pair states, the flat-band states exist between two nodal points that have opposite topological charges 
and the number of the flat-band states is two for each edge. 
On the other hand, in the cases of same-sign pair states [Fig.~\ref{fig:band}(d)], the flat-band states exist at $k_y=0$, 
and the number of the flat-band states across $k_y=0$ and between two nodal points in positive $k_y$ is four and two for each edge, respectively.
In these cases, the number of flat-band states has a one to one correspondence with $\left|N(k_y)\right|$, which is shown in Fig.~\ref{fig:windingABS}.
Note that in Fig.~\ref{fig:band}(d) the finite-size effect creates a gap at $k_y=0$.
We have confirmed that there is no gap at $k_y=0$ by using the recursive Green's function method (see Appendix~\ref{appendix:greenfunc}). In addition to the flat-band edge states that appear when nodes exist in the bulk spectrum, note that we find  edge states within the gap, although not at zero energy, even in the fully gapped case. These can be attributed to sign changes in the gap that still appear in a fully gapped $d_{xy}$ superconductor.

\begin{figure}
\begin{center}
   \includegraphics[width=\linewidth]{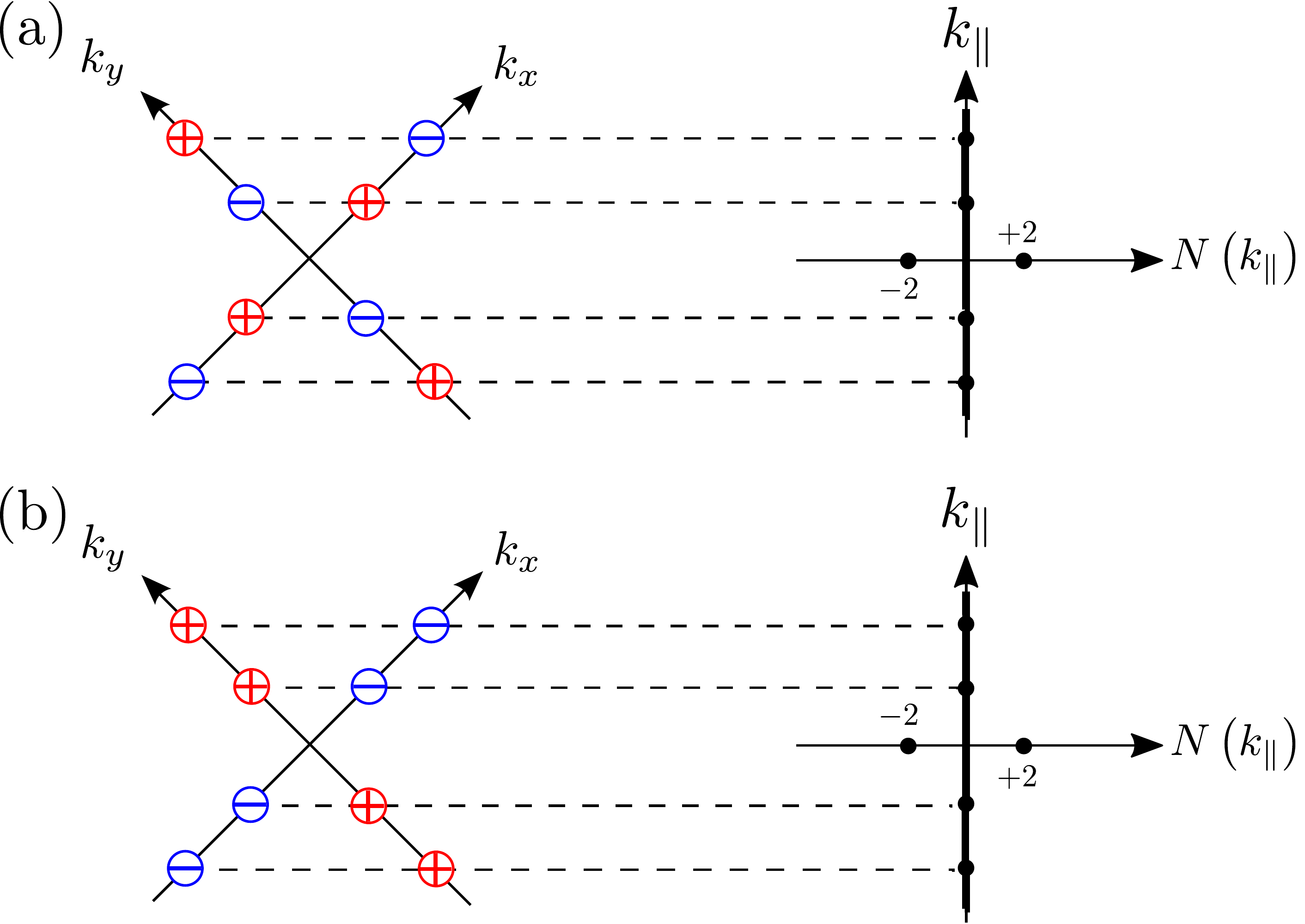}
   \caption{Schematic pictures of the relation between $W_{\mathcal{L}}$ (left) and $N(k_\parallel)$ (right) in the case of (a) opposite-sign pair and (b) same-sign pair states. We consider the edges running along the $(1,1)$ direction.
   Red and blue points indicate $W_{\mathcal{L}}=+2$ and $-2$, respectively. }
 \label{fig:windingABS11}
 \end{center}
\end{figure}

In actual experiments misalignments would appear, and it is worth  mentioning the consequences of this on the distinct topological phases and the resultant anisotropy of the number of Andeev flat bound states.
The one-to-one correspondence between the number of flat-band states and $\left|N\left(k_\parallel\right)\right|$ is also useful for the edge in  other directions. 
For instance, consider the edges running along the $(1,1)$ direction and denote the wave-vector component $k_\parallel$ parallel to the edges. 
Figures~\ref{fig:windingABS11}(a) and \ref{fig:windingABS11}(b) show the 1D winding number $\left|N\left(k_\parallel\right)\right|$ and the topological charge $W_{\mathcal{L}}$ for the cases of opposite-sign pair and same-sign pair states, respectively.
For both cases $\left|N\left(k_\parallel\right)\right|=0$ for any $k_\parallel$; therefore, there are no Andreev flat-band states.

Finally, we note that the examination of the Andreev bound state spectra should take into account interaction effects. It has been pointed out that due to the large density of states intrinsic to flat-bands, they are susceptible to surface instabilities~\cite{Schnyder-11,Potter-14}. The most likely candidate is edge ferromagnetism that spits the flat-bands~\cite{Potter-14}.  Such a surface instability is seen in tunneling spectroscopy experiments on the cuprate superconductor YBa$_2$CU$_3$O$_7$ where the zero-bias conductance peak is seen to split into two below an edge transition temperature that is approximately $0.1 T_c$~\cite{Covington-97}. We leave the study of possible edge instabilities of Andreev flat-band states due to interactions in the context of the models examined here to future work. 


\section{Conclusion}
\label{sec:Conclusion}
We have studied nodal topological charges in $d$-wave superconducting monolayer FeSe to help understand the origin of a fully gapped  $d$-wave state.
The nodal points that arise when interband spin-orbit coupling is sufficiently strong have $2\mathbb{Z}$ topological charges that give rise to zero-energy dispersionless Andreev edge bound states. The momentum space distribution of the nodal charges depends strongly on the orbital character of the superconducting state, allowing this  to  be probed through the observation of Andreev bound states.

\section*{Acknowledgments}
We thank Philip Brydon, Andrey Chubukov, Hirokazu Tsunetsugu, and Michael Weinert for useful discussions. 
Our numerical calculations were partly carried out at the Supercomputer Center, The Institute for Solid State Physics,The University of Tokyo.
T. N was supported by Japan Society 433 for the Promotion of Science  through Program for Leading Graduate Schools (MERIT).

\vspace{5mm}
\clearpage
\appendix

\section{lattice model}\label{appendix:latticemodel}
In order to obtain the lattice model which corresponds to Eq.~(\ref{eq:Hbdg}), we replace $k_i\to \sin k_i$ and $\left(k_x^2+k_y^2\right)/(2m) \to -2t \left(\cos k_x +\cos k_y\right)+4t$ where $t^{-1}=2m$ in  Eq.~(\ref{eq:Hbdg}) (the lattice constant is unity).
We use $A_{\bm{i}\sigma}$ and $B_{\bm{i}\sigma}$, which are annihilation operators of two orbital, spin $\sigma=\up$ and $\down$ electrons at $\bm{i}$,  and we divide $\mathcal{H}$ into $\mathcal{H}_0$,  $\mathcal{H}_{\rm SOC}$, and $\mathcal{H}_{\rm \Delta}$.
They are given by
\begin{widetext}
\begin{eqnarray}
\mathcal{H}_0 &=&-t \sum_{\left<\bm{i}, \bm{j}\right>,\sigma}\left[A^\dag_{\bm{i}\sigma}A_{\bm{j}\sigma}+B^\dag_{\bm{i}\sigma}B_{\bm{j}\sigma}\right]
-\left(\mu-4t\right)\sum_{\bm{i},\sigma}\left[A^\dag_{\bm{i}\sigma}A_{\bm{i}\sigma}+B^\dag_{\bm{i}\sigma}B_{\bm{i}\sigma}\right] \nonumber\\
&&+\frac{a}{4}\sum_{\bm{i},\sigma}\left[A^\dag_{\bm{i}\sigma}A_{\bm{i}+\bm{x}+\bm{y}\sigma}+A^\dag_{\bm{i}+\bm{x}+\bm{y}\sigma}A_{\bm{i}\sigma}-\left(A^\dag_{\bm{i}\sigma}A_{\bm{i}+\bm{x}-\bm{y}\sigma}+A^\dag_{\bm{i}+\bm{x}-\bm{y}\sigma}A_{\bm{i}\sigma}\right)\right] \nonumber\\
&&-\frac{a}{4}\sum_{\bm{i},\sigma}\left[B^\dag_{\bm{i}\sigma}B_{\bm{i}+\bm{x}+\bm{y}\sigma}+B^\dag_{\bm{i}+\bm{x}+\bm{y}\sigma}B_{\bm{i}\sigma}-\left(B^\dag_{\bm{i}\sigma}B_{\bm{i}+\bm{x}-\bm{y}\sigma}+B^\dag_{\bm{i}+\bm{x}-\bm{y}\sigma}B_{\bm{i}\sigma}\right)\right], \\
\mathcal{H}_{\rm SOC} &=&-\frac{v_{\rm so}}{2}\sum_{\bm{i}}\left[\left\{A^\dag_{\bm{i}\up}B_{\bm{i}+\bm{x}\down}-A^\dag_{\bm{i}+\bm{x}\up}B_{\bm{i}\down}\right\}-\left\{A^\dag_{\bm{i}\down}B_{\bm{i}+\bm{x}\up}-A^\dag_{\bm{i}+\bm{x}\down}B_{\bm{i}\up}\right\} \right. \nonumber\\
&&\quad\quad\quad\quad+\left.\left\{B^\dag_{\bm{i}\up}A_{\bm{i}+\bm{x}\down}-B^\dag_{\bm{i}+\bm{x}\up}A_{\bm{i}\down}\right\}-\left\{B^\dag_{\bm{i}\down}A_{\bm{i}+\bm{x}\up}-B^\dag_{\bm{i}+\bm{x}\down}A_{\bm{i}\up}\right\} \right]\nonumber \\
&&+\frac{v_{\rm so}}{2i}\sum_{\bm{i}} \left[\left\{A^\dag_{\bm{i}\up}B_{\bm{i}+\bm{y}\down}-A^\dag_{\bm{i}+\bm{y}\up}B_{\bm{i}\down}\right\}+\left\{A^\dag_{\bm{i}\down}B_{\bm{i}+\bm{y}\up}-A^\dag_{\bm{i}+\bm{y}\down}B_{\bm{i}\up}\right\} \right. \nonumber\\
&&\quad\quad\quad\quad+\left.\left\{B^\dag_{\bm{i}\up}A_{\bm{i}+\bm{y}\down}-B^\dag_{\bm{i}+\bm{y}\up}A_{\bm{i}\down}\right\}+\left\{B^\dag_{\bm{i}\down}A_{\bm{i}+\bm{y}\up}-B^\dag_{\bm{i}+\bm{y}\down}A_{\bm{i}\up}\right\} \right], \\
\mathcal{H}_{\rm \Delta}&=&-\frac{\Delta_2}{4k_0^2}\sum_{\bm{i}}\left[A_{\bm{i}\up}^\dag A_{\bm{i-\bm{x}-\bm{y}}\down}^\dag+A_{\bm{i}\up}^\dag A_{\bm{i+\bm{x}+\bm{y}}\down}^\dag-\left(A_{\bm{i}\up}^\dag A_{\bm{i-\bm{x}+\bm{y}}\down}^\dag+A_{\bm{i}\up}^\dag A_{\bm{i+\bm{x}-\bm{y}}\down}^\dag\right)\right.\nonumber  \\
&&\quad\quad\quad\quad\left.-\left\{A_{\bm{i}\down}^\dag A_{\bm{i-\bm{x}-\bm{y}}\up}^\dag+A_{\bm{i}\down}^\dag A_{\bm{i+\bm{x}+\bm{y}}\up}^\dag-\left(A_{\bm{i}\down}^\dag A_{\bm{i-\bm{x}+\bm{y}}\up}^\dag+A_{\bm{i}\down}^\dag A_{\bm{i+\bm{x}-\bm{y}}\up}^\dag\right)\right\}\right] \nonumber \\
&&-\frac{\Delta_2}{4k_0^2}\sum_{\bm{i}}\left[B_{\bm{i}\up}^\dag B_{\bm{i-\bm{x}-\bm{y}}\down}^\dag+B_{\bm{i}\up}^\dag B_{\bm{i+\bm{x}+\bm{y}}\down}^\dag-\left(B_{\bm{i}\up}^\dag B_{\bm{i-\bm{x}+\bm{y}}\down}^\dag+B_{\bm{i}\up}^\dag B_{\bm{i+\bm{x}-\bm{y}}\down}^\dag\right)\right.\nonumber  \\
&&\quad\quad\quad\quad\left.-\left\{B_{\bm{i}\down}^\dag B_{\bm{i-\bm{x}-\bm{y}}\up}^\dag+B_{\bm{i}\down}^\dag B_{\bm{i+\bm{x}+\bm{y}}\up}^\dag-\left(B_{\bm{i}\down}^\dag B_{\bm{i-\bm{x}+\bm{y}}\up}^\dag+B_{\bm{i}\down}^\dag B_{\bm{i+\bm{x}-\bm{y}}\up}^\dag\right)\right\}\right] \nonumber \\
&&+\Delta_0\sum_{\bm{i}}\left[A_{\bm{i}\up}^\dag A_{\bm{i}\down}^\dag-A_{\bm{i}\down}^\dag A_{\bm{i}\up}^\dag-\left(B_{\bm{i}\up}^\dag B_{\bm{i}\down}^\dag-B_{\bm{i}\down}^\dag B_{\bm{i}\up}^\dag\right) \right] \nonumber \\
&+&{\rm H.c.}
\end{eqnarray}
\end{widetext}
\section{Energy spectrum using the Green's function method}
\label{appendix:greenfunc}
Our Hamiltonian matrix of the edge problem has a simple band form, 
\begin{eqnarray}
\mathcal{H}=\left( \begin{array}{cccccccc}
A & B & 0 & 0 & 0 & 0 & \cdot  & \cdot \\
B^\dagger & A & B & 0 & 0 & 0 & \cdot & \cdot \\
0 & B^\dagger & A & B & 0 & 0 & \cdot & \cdot \\
0 & 0 & B^\dagger & A & B & 0  \\
\cdot & \cdot & \cdot  \\
\cdot & \cdot & \cdot  
\end{array}
\right), 
\end{eqnarray}
where $A$ and $B$ are small square matrices of order 8 (or 4 in the reduced block form).  
L\'opez Sancho \textit{et al.}\cite{Sancho-84} developed a highly convergent iterative scheme to calculate the surface and bulk Green's functions ($G_{00}$ and $G_{\infty\infty}$, respectively) for this form of Hamiltonian.   At the $i$th iteration, the (renormalized) $G_{00}$ is given in terms of effective interaction with the $2^i$th layer: 
\begin{eqnarray}
(\omega I - \epsilon_i^\mathrm{s} ) G_{00} = I + \alpha_i G_{2^i,0} 
\end{eqnarray}
and other elements are given by
\begin{eqnarray}
(\omega I - \epsilon_i ) G_{2^in,0} &=& \beta_i G_{2^i(n-1),0} + \alpha_i G_{2^i(n+1),0} , \\
(\omega I - \epsilon_i ) G_{2^in,2^in} &=& I + \beta_i G_{2^i(n-1),2^in} + \alpha_i G_{2^i(n+1),2^in}, \nonumber \\
\end{eqnarray}
where $\omega$ is an energy with a small imaginary part $i\eta$ and ($\omega$-dependent) energy matrices $\epsilon_i^\mathrm{s}$, $\epsilon_i$, $\alpha_i$, and $\beta_i$ are determined recursively starting from $\epsilon_0^\mathrm{s}=\epsilon_0=A$, $\alpha_0=B$, and $\beta_0=B^\dagger$.   As the iteration proceeds, the effective interactions $\alpha_i$ and $\beta_i$ decay quickly.  
We take $\eta/t=10^{-5}$, and the iteration is truncated when $|\alpha_i /t |, |\beta_i /t |< 10^{-7}$.  The required number of iterations is at most 20.   

Figure~\ref{fig:G_edge} shows $k_y$-resolved spectral functions obtained with this method, 
\begin{eqnarray}
N_n (k_y, E) = - \frac{1}{\pi} \  \mathrm{Im} \ \mathrm{Tr} \ G_{nn} (k_y, E+ i \eta),
\end{eqnarray}
with $n=0$ (edge) and $n=\infty$ (bulk), for the four parameter sets used in Figs.~\ref{fig:band}(a)-\ref{fig:band}(d). A blowup of spectral functions near $k_y\sim0$ is shown in Fig.~\ref{fig:blowup}.

\begin{widetext}

\begin{figure}[h]
\includegraphics[width=1.0\columnwidth]{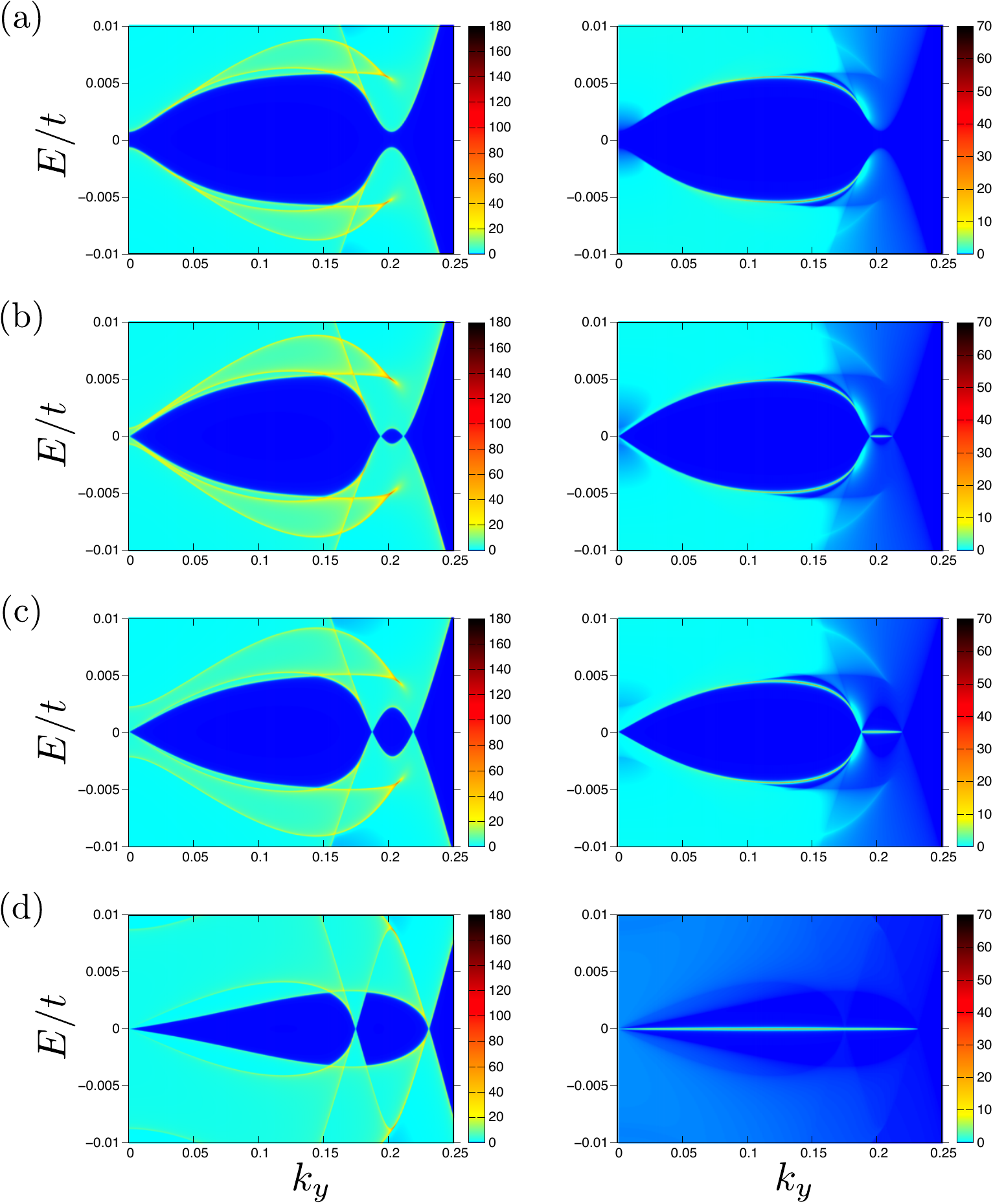}
\caption{\label{fig:G_edge} 
Momentum-resolved spectral function calculated by the Green's function method.  
Left (right) panels provide the local density of states at the bulk (edge).  
The dark blue area represents a no-state region.  
(a) Full gap, (b) and (c) opposite-sign pairs of nodal points, and (d) same-sign pair of nodal points.  The energy is given in units of $t$. 
}
\end{figure}

\begin{figure}[h]
\includegraphics[width=1.0\columnwidth]{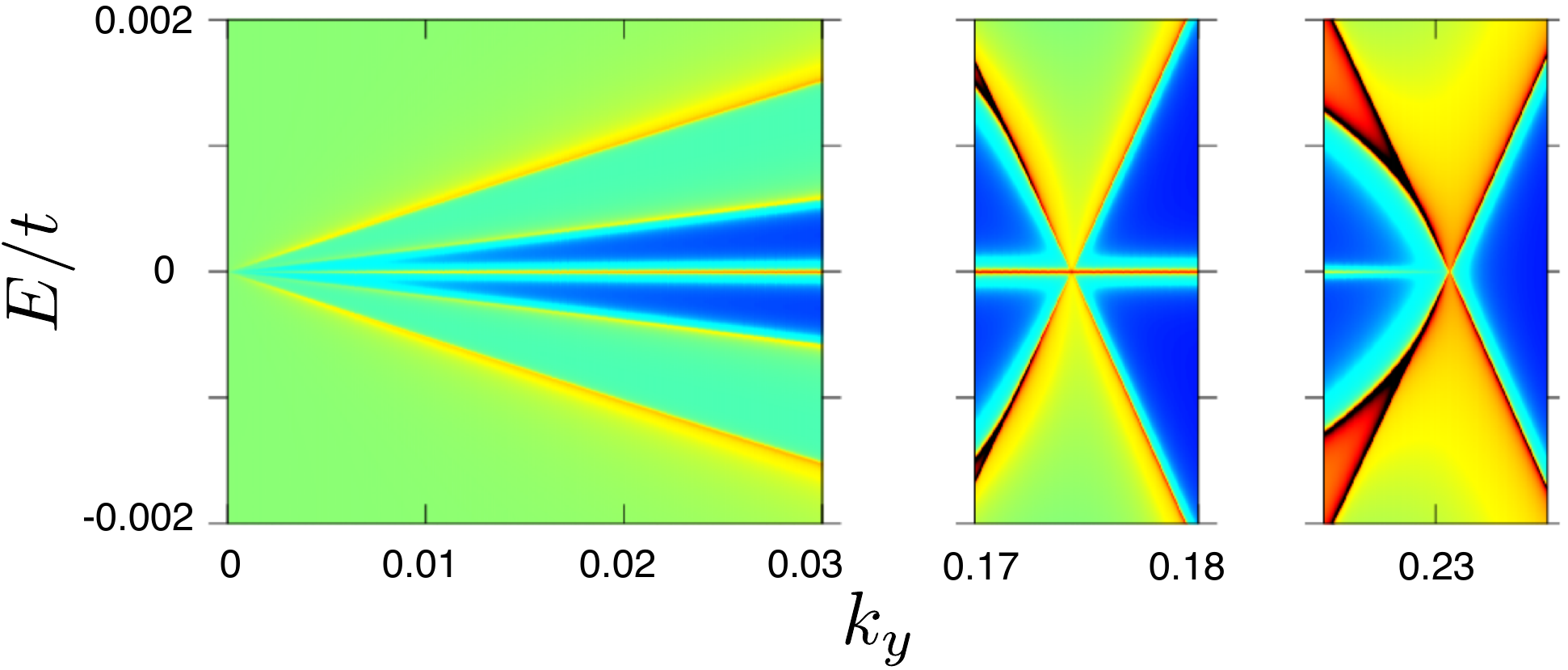}
\caption{\label{fig:blowup} 
Blowup of the spectral function (edge+bulk) of the parameter set in Fig.~\ref{fig:band}(d) around gapless regions  
with fine resolution in energy and momentum. }
\end{figure}

\clearpage
\end{widetext}

\end{document}